\documentclass[%
 reprint,
 superscriptaddress,
showpacs,
nofootinbib,
 amsmath,amssymb,
 aps,
 prc,
]{revtex4-1}

\usepackage[latin1]{inputenc}
\usepackage{graphicx}
\usepackage{amssymb}
\usepackage{amsmath}
\usepackage{multirow}

\newcommand{\nuc}[2]{\ensuremath{^{#1}}#2}
\newcommand{\sys}[4]{\ensuremath{^{#1}}#2+\ensuremath{^{#3}}#4}
\newcommand{\AM}{$A\,$MeV}
\begin{document}

\title{New isospin effects in central heavy-ion collisions at Fermi energies}

\author{F.~Gagnon-Moisan}
\altaffiliation{Current address: PTB Brauncshweig, Bundesallee 100, 
38116 Braunschweig, Germany}
\affiliation{Institut de Physique Nucléaire, CNRS-IN2P3 and Université
Paris-Sud 11, F-91406 Orsay France}
\affiliation{Universit\'e Laval, Québec, G1V 0A6 Canada}
\author{E.~Galichet}
\affiliation{Institut de Physique Nucléaire, CNRS-IN2P3 and Université
Paris-Sud 11, F-91406 Orsay France}
\affiliation{Conservatoire National des Arts et M\'etiers, F-75141 Paris
Cedex 03, France}
\author{M.-F.~Rivet}
\email[]{rivet@ipno.in2p3.fr}
\author{B.~Borderie}
\affiliation{Institut de Physique Nucléaire, CNRS-IN2P3 and Université
Paris-Sud 11, F-91406 Orsay France}
\author{M.~Colonna}
\affiliation{Laboratori Nazionali del Sud-Istituto Nazionale Fisica Nucleare, 
I-95123 Catania, Italy}
\author{R.~Roy}
\affiliation{Universit\'e Laval, Québec, G1V 0A6 Canada}
\author{G.~Ademard}
\affiliation{GANIL, CEA-DSM/CNRS-IN2P3, F-14076 Caen Cedex, France}
\affiliation{Institut de Physique Nucléaire, CNRS-IN2P3 and Université
Paris-Sud 11, F-91406 Orsay France}
\author{M.~Boisjoli}
\affiliation{Universit\'e Laval, Québec, G1V 0A6 Canada}
\affiliation{GANIL, CEA-DSM/CNRS-IN2P3, F-14076 Caen Cedex, France}
\author{E.~Bonnet},
\affiliation{GANIL, CEA-DSM/CNRS-IN2P3, F-14076 Caen Cedex, France}
\author{R.~Bougault}
\affiliation{LPC Caen, ENSICAEN, Universit\'e de Caen, CNRS-IN2P3, F-14050
Caen Cedex, France}
\author{A.~Chbihi},
\author{J.D.~Frankland},
\affiliation{GANIL, CEA-DSM/CNRS-IN2P3, F-14076 Caen Cedex, France}
\author{D.~Guinet}
\author{P.~Lautesse}
\altaffiliation{Present address: S2HEP (EA4148), UCBL/ENSL, Université de Lyon, 
Villeurbanne, France}
\affiliation{Institut de Physique Nucl\'eaire, UCBL, Universit\'e de Lyon, 
CNRS-IN2P3, F-69622 Villeurbanne Cedex, France}
\author{E.~Legou\'ee}
\author{N.~Le~Neindre}
\affiliation{LPC Caen, ENSICAEN, Universit\'e de Caen, CNRS-IN2P3, F-14050
Caen Cedex, France}
\author{L.~Manduci}
\affiliation{EAMEA, CC19 50115 Cherbourg-Octeville Cedex, France}
\affiliation{GANIL, CEA-DSM/CNRS-IN2P3, F-14076 Caen Cedex, France}
\author{P.~Marini},
\affiliation{GANIL, CEA-DSM/CNRS-IN2P3, F-14076 Caen Cedex, France}
\author{P.~Napolitani}
\affiliation{Institut de Physique Nucléaire, CNRS-IN2P3 and Université
Paris-Sud 11, F-91406 Orsay France}
\author{M.~P\^arlog}
\affiliation{LPC Caen, ENSICAEN, Universit\'e de Caen, CNRS-IN2P3, F-14050
Caen Cedex, France}
\affiliation{National Institute for Physics and Nuclear Engineering,
RO-76900 Bucharest-M\u{a}gurele, Romania}
\author{P.~Paw{\l}owski}
\affiliation{IFJ-PAN, 31-342 Krak\'ow, Poland}
\author{E.~Rosato}
\author{M.~Vigilante}
\affiliation{Dipartimento di Scienze Fisiche e Sezione INFN, Universit\`a
di Napoli ''Federico II'', I-80126 Napoli, Italy}
\collaboration{INDRA collaboration}
\homepage{http://indra.in2p3.fr/spip/}
\noaffiliation

\date{\today}

\begin{abstract}
Isospin effects on multifragmentation properties were studied thanks to
nuclear collisions between different isotopes of xenon beams and tin targets.
It is shown that, in central collisions leading to multifragmentation, 
the mean number of fragments and their
mean kinetic energy  increase with the neutron-richness of the total system. 
Comparisons with a stochastic transport model allow to attribute the 
multiplicity increase to the multifragmentation stage, before secondary decay.
The total charge bound in fragments is proposed as an alternate variable to
quantify preequilibrium emission  and to investigate symmetry energy effects.
\end{abstract}


\maketitle

\section{\label{sec:intro}Introduction}

One of the present motivations for investigating heavy-ion collisions at
intermediate energy is the improvement of the knowledge of the Equation 
of State (EOS) for nuclear matter. More specifically the formulation of 
an adequate symmetry term is required to progress, and the density 
dependence of this term both at sub- and  supra-normal density is still 
debated. Further experimental constraints are clearly necessary. 
Within the next decade physicists expect the advent of new heavy-ion 
accelerators, providing high-intensity exotic beams, in order to study 
reactions covering a broad range of isospin ($N/Z$) ratios; in the meantime 
information on isospin effects can be obtained thanks to a judicious choice of 
projectile-target couples. 
Comparisons of several calculated and experimentally measured 
isospin-dependent variables already provided some hints on the symmetry term, 
but the results are still highly model- and 
experiment-dependent~\cite{Bar05,Bao08,*[{}] [{ and references therein}]
Riv09,Tsa04,Che05,Zha08,Tsa09}.

In this line the INDRA collaboration has studied collisions between 
\nuc{124,136}{Xe} projectiles, with an incoming energy of 32 and 45~\AM{}, 
and \nuc{112,124}{Sn} targets.  
 We show in this paper that, in central collisions, the mean charged-product 
and fragment (Z$\geq$5) multiplicities, and mean fragment kinetic energies,
depend on the isospin of the total system.
The commonly accepted reaction scenario for central heavy-ion collisions
around Fermi energy, validated by stochastic transport models, is the
following~\cite{Bor08}:
there is firstly  a compression phase, the strength of
which depends on the masses and mass asymmetry of the incident partners.
This stage is followed by an expansion phase accompanied by the emission of
fast preequilibrium particles. The diluted system enters the spinodal region
of the phase diagram and eventually breaks into
several excited fragments, light particles and nucleons (multifragmentation); 
when the configuration is frozen, namely when the nuclear interaction 
between fragments becomes negligible, one speaks of the ``freeze-out'' stage.
Finally the charged products move apart, further accelerated by the Coulomb
force, while losing their excitation energy through evaporation. 
 
 In this framework we shall test, through comparisons with a stochastic 
 transport model, the origin of the  isospin dependence observed for the 
 aforementioned variables.

\section{\label{sec:exp}Experiment} 

Beams of 32 and 45~\AM{} \nuc{124,136}Xe accelerated by the 
\textit{Grand Acc\'el\'erateur National d'Ions Lourds} (GANIL) impinged on
530~$\mu$g.cm$^{-2}$ \nuc{112,124}Sn targets. Beam intensity was about 3-5
$\times 10^{7}$ particles per second to avoid event pile-up.
Charged products were detected and identified with the $4\pi$ INDRA 
array~\cite{I3-Pou95}, which comprises 336 two- or three-member telescopes
arranged in a cylindrical geometry around the beam axis. The array was
upgraded as follows~\cite{T47Gag10}:  12 (300 $\mu$m)silicon-(14 cm)CsI(Tl) 
telescopes replaced the phoswiches of the first ring, and 7 of the 300 $\mu$m 
silicon  wafers were replaced by 150 $\mu$m wafers, with an increased amplifier 
gain, in order to identify isotopes up to nitrogen. In the standard
silicon-CsI(Tl) telescopes (2-45$^o$) elements were identified within one
charge unit; in addition between 14$^o$ and 45$^o$ these telescopes provided
isotopic identification up to carbon. 
The same Z-resolution was obtained in the first stage ionization chamber-silicon 
or ionization chamber-CsI(Tl)) telescopes up to Z=20. For low energy heavier nuclei, 
Z are known with an uncertainty of 1-2 charge units~\cite{T43Moi08}. 
In this paper, the term ``fragments'' refers to charged products with a charge 
$Z\geqslant5$ whereas light charged particles (lcp) stands for particles with 
a charge $Z\leqslant2$. 

Most of data taking was done with an on-line trigger requiring at least 
four fired INDRA telescopes. For a better overview of the reactions,  
some inclusive measurements (one telescope fired) were also performed. 

Absolute cross sections were derived from the measured target thicknesses,
the counting of ions collected in the Faraday cup located at the end of the 
beam line and the acquisition dead time. The charge of ions reaching the cup 
was obtained using the formulae of reference~\cite{Schi01}. 
The error on cross sections is estimated to
15\%, while relative errors between two systems are lower, around  3\%.

Table~\ref{tab:Sys} summarizes the studied systems and the isospin contents
of the total systems. Note that two of the systems have
the same N/Z, allowing to study entrance channel mass-asymmetry effects.  
We also indicate the measured inclusive cross sections (trigger multiplicity
$\geq$1). In the off-line analysis we required at least one identified 
charged product and rejected events formed of a single charged product 
with an atomic number close to the projectile one, in order to exclude 
elastic scattering. The cross sections so obtained are close to the
reaction cross sections calculated with the systematics of 
Kox~\cite{Kox84} at 32~\AM{}
whereas they are smaller by about 1 barn at 45~\AM{}.

\begin{table}[h]
\begin{tabular}{cclccccc}
Projectile & Target & $(N/Z)$ &$E_{proj.}$& $\sigma$ &$E_{proj.}$& $\sigma$  \\
& & syst.& (\AM{}) & (b) & (\AM{}) & (b) \\
\hline
$^{124}Xe$ & $^{112}Sn$ & 1.269 & 32 & 5.16 & 45 & 4.73  \\
$^{124}Xe$ & $^{124}Sn$ & 1.385 & 32 & 5.14 & -   &  - \\
$^{136}Xe$ & $^{112}Sn$ & 1.385 & 32 & 5.90 & 45 &  - \\
$^{136}Xe$ & $^{124}Sn$ & 1.50  & 32 & 5.81 & 45 &  5.02 \\
\end{tabular}
\caption{\label{tab:Sys}Available systems and measured cross
sections. Systematic errors on cross sections are estimated to 15\%}
\end{table}

\begin{figure}
\includegraphics[width=0.5\textwidth]{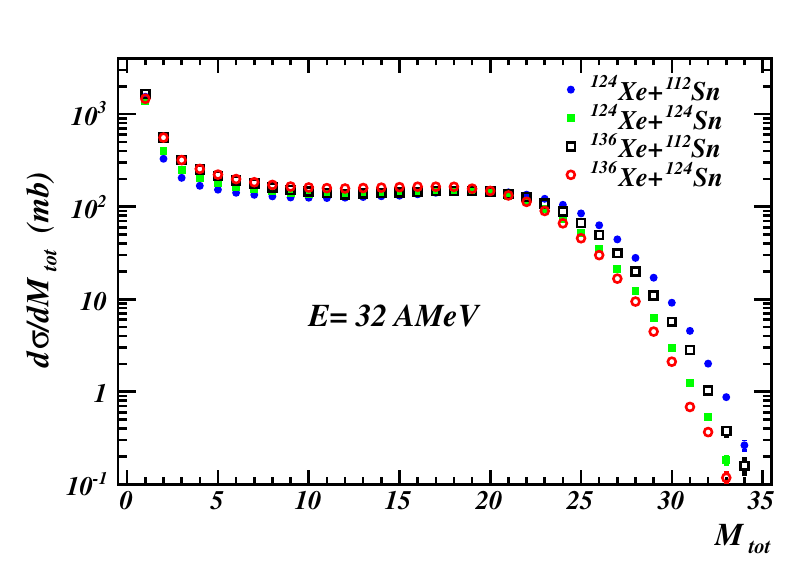}\\
\caption{(color on line) Charged-product multiplicity distributions for 
the different systems at 32~\AM{}. Error bars are statistical.} \label{fig:mtot}
\end{figure}

Figure~\ref{fig:mtot} shows the charged-product multiplicity distributions of
identified nuclei at 32~\AM{}, for data with an on-line trigger multiplicity
$M\geq$1.  For low multiplicities the cross sections 
are higher for the neutron-rich system; the trend is reversed for 
multiplicities larger than 20,
and the distribution extends to higher values for the neutron-poor system.
The two systems \sys{124}{Xe}{124}{Sn} and \sys{136}{Xe}{112}{Sn}
display intermediate values.
A first isospin effect appears already on raw variables: higher
charged products multiplicities are observed when the system contains less
neutrons; it is indeed expected that a neutron-rich system will
preferentially emit neutrons to the detriment of light charged particles,
which dominate in the charged product multiplicity.

\section{\label{sec:csev}Compact shape events from central collisions}

In the following we shall report on data obtained with an on-line trigger
$M\geq$4 for which we got a very high statistics.
As in previous studies~\cite{I28-Fra01,I39-Hud03,I40-Tab03,I57-Tab05}
we select quasi-complete events by requiring that the 
sum of the charges of the detected products, $Z_{tot}$, be at least equal to 80. 
Due to the response of INDRA to the kinematics of quasi-symmetric 
collisions such as Xe+Sn, this choice favourizes \textit{de facto}  
central collisions.
We then isolate compact shape events (quasi-fusion) through the 
additional condition that the flow angle ($\theta_{flow}$) be greater than 
$60^{\circ}$. Let us recall that $\theta_{flow}$ characterizes the main
direction of matter emission in the reaction center of mass 
and is determined by the kinetic energy flow-tensor calculated from fragment
momenta. More than 10$^5$ events were selected in all cases, corresponding to 
measured cross-sections of 30-40~mb at 32~\AM{}, and 20-25~mb at 45~\AM{}. 
These values are slightly larger than those previously measured for the 
\sys{129}{Xe}{nat}{Sn} system~\cite{I40-Tab03}. The total cross-section for 
quasi-fusion, taking into account detection efficiency and selection biases, 
is estimated to reach  $\sim$200-250~mb at 32~\AM{}, 
and $\sim$170~mb at 45~\AM{} (the difference between estimated and 
measured cross-sections is due to the limits on $\theta_{flow}$ and 
on $Z_{tot}$ imposed in the selection, which cut part of events that can be
attributed to the quasi-fusion mechanism~\cite{T16Sal97}).

\subsection{Multiplicities}
\begin{table*}[hbt] 
\caption{Average measured multiplicities of charged products, 
$M_{tot}$, of light
charged particles (H, He), $M_{lcp}$,  and of fragments, $M_{frag}$ for the
different systems studied. Number in parentheses are the standard deviations
of the corresponding distributions. Statistical errors on mean values are
smaller than 0.01 for $M_{tot}$ and $M_{lcp}$ and $M_{frag}$.
Results on line 3 are from~\cite{I40-Tab03}.}
\label{tab:Multcent}
\begin{tabular}{p{23.5mm}p{22mm}p{22mm}p{22mm}|p{22mm}p{22mm}p{22mm}}
& \multicolumn{3}{c|}{E/A=32 MeV} & \multicolumn{3}{c}{E/A=45 MeV} \\ \hline
System & $\langle M_{tot}\rangle$ & $\langle M_{lcp}\rangle$ & $\langle M_{frag}\rangle$ &
$\langle M_{tot}\rangle$ & $\langle M_{lcp}\rangle$ & $\langle M_{frag}\rangle$ \\ \hline
\sys{124}{Xe}{112}{Sn} & 25.12 (2.90) & 19.66 (3.24) & 4.11 (1.16) & 
32.65 (3.24) & 25.96 (3.68) & 4.30 (1.18) \\  
\sys{124}{Xe}{124}{Sn} & 23.71 (2.88) & 18.06 (3.22) & 4.23 (1.17) & - & - & - \\
\sys{129}{Xe}{nat}{Sn} & 23.92 (3.00) & 18.37 (3.28) & 4.13 (1.17) &
31.4 (3.21) & 24.57 (3.64) & 4.39 (1.20) \\  
\sys{136}{Xe}{112}{Sn} & 24.23 (3.01) & 18.38 (3.30) & 4.36 (1.18) & 
31.04 (3.28) & 24.02 (3.67) & 4.52 (1.20) \\  
\sys{136}{Xe}{124}{Sn} & 23.07 (3.00) & 16.97 (3.28) & 4.54 (1.20) & 
30.0 (3.26) & 22.68 (3.66) & 4.71 (1.23) \\
\end{tabular}
\end{table*}

\begin{figure}
\includegraphics[width=0.95\columnwidth]{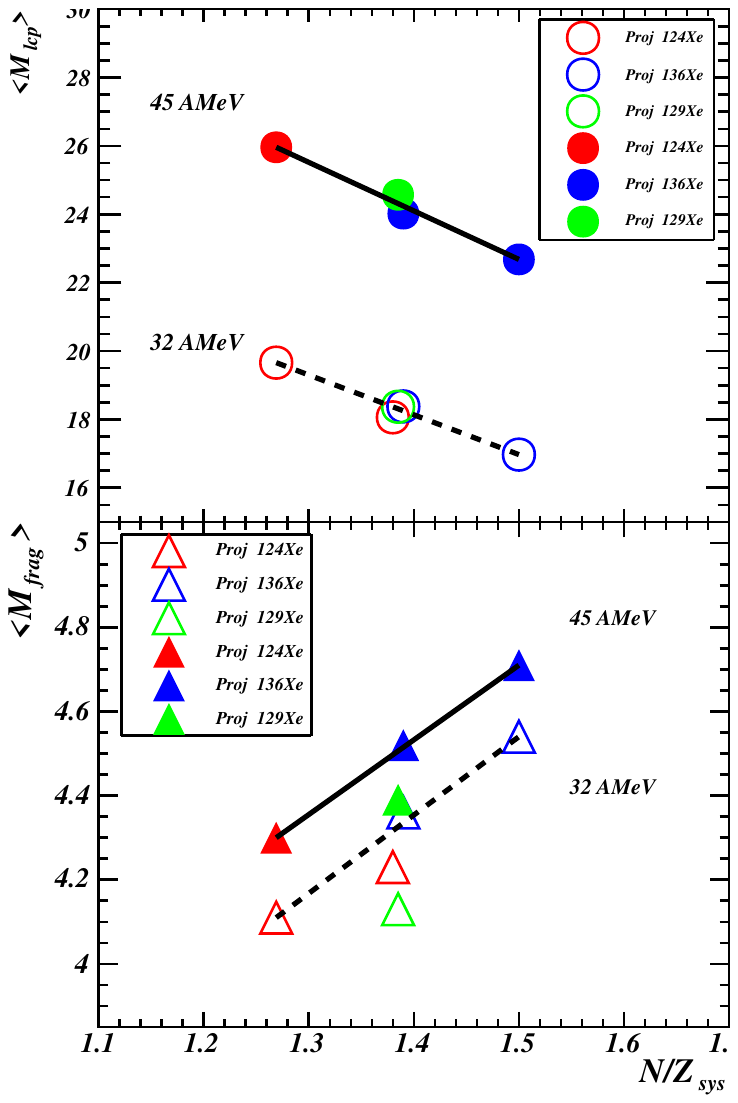}
\caption{(color on line) Evolution of the light charged particle and fragment 
multiplicities
versus the N/Z of the total systems, at the two energies. Error bars are
smaller than the size of the symbols.}\label{fig:lcpfrag}
\end{figure}

Table~\ref{tab:Multcent} displays the different average multiplicities measured
for the selected class of events, for all studied systems: $M_{tot}$ is the 
total charged product multiplicity, $M_{lcp}$ and $M_{frag}$ refer to the 
light charged particle and fragment multiplicities, which are shown in
fig.~\ref{fig:lcpfrag}. Results from
the reaction \sys{129}{Xe}{nat}{Sn} have been added~\cite{I40-Tab03}: 
the composite system is the same as those formed by  \sys{136}{Xe}{112}{Sn} and 
\sys{124}{Xe}{124}{Sn} (the average mass number of \nuc{nat}{Sn} is 119). 
As in the unsorted data, at each
beam energy, the average multiplicities of charged products and of light 
charged particles decrease for increasing N/Z of the total system.
Conversely the fragment multiplicity increases. The three systems with the
same N/Z show very close mean lcp multiplicity values, which
indicates that the entrance channel mass asymmetry has a small influence.
The same is not true for mean fragment multiplicities, whereas their standard
deviations are equal. While we may suspect some difference in the INDRA
response between the present data and those of~\cite{I40-Tab03}, such an explanation
does not hold for \sys{124}{Xe}{124}{Sn} and \sys{136}{Xe}{112}{Sn} at 32
\AM{}. In that case we might envisage some physical reason, which requires
further investigation.
Previous studies~\cite{Dem96,Kun96}  already indicated that more
fragments are emitted for neutron-rich systems. 
The available energy per nucleon, at the same incident energy, is the
same for all systems, within 1\%, so the increased number of fragments 
cannot be explained by a larger available energy. Moreover $M_{frag}$ 
modestly increases between 32 and 45~\AM{}: it was indeed shown
in~\cite{I40-Tab03} that for the system \sys{129}{Xe}{nat}{Sn} $M_{frag}$
presents a maximum for an incident energy around 40~\AM{}.

 In previous works the difference in fragment multiplicities was attributed 
 to phase-space effects~\cite{Kun96}, because statistical calculations 
 (EES~\cite{Fri90}) reproduce the observation, 
or to sequential decay effects~\cite{Dem96}. In the present data
the fragment multiplicity increases by 10\%, both at 32 and 45~\AM{},
between the lightest and the heaviest systems. This corresponds
to the mass increase between these systems, which may recall the 
scaling law observed in~\cite{I12-Riv98},  expected if 
multifragmentation originates from volume instabilities (spinodal decomposition).
Using stochastic transport codes we can test whether the increased fragment 
multiplicity arises from the dynamical or from the secondary decay stage
of the reaction.

\subsection{Fragment kinetic energies}

\begin{figure*}[htb]
\includegraphics[width=0.45\textwidth]{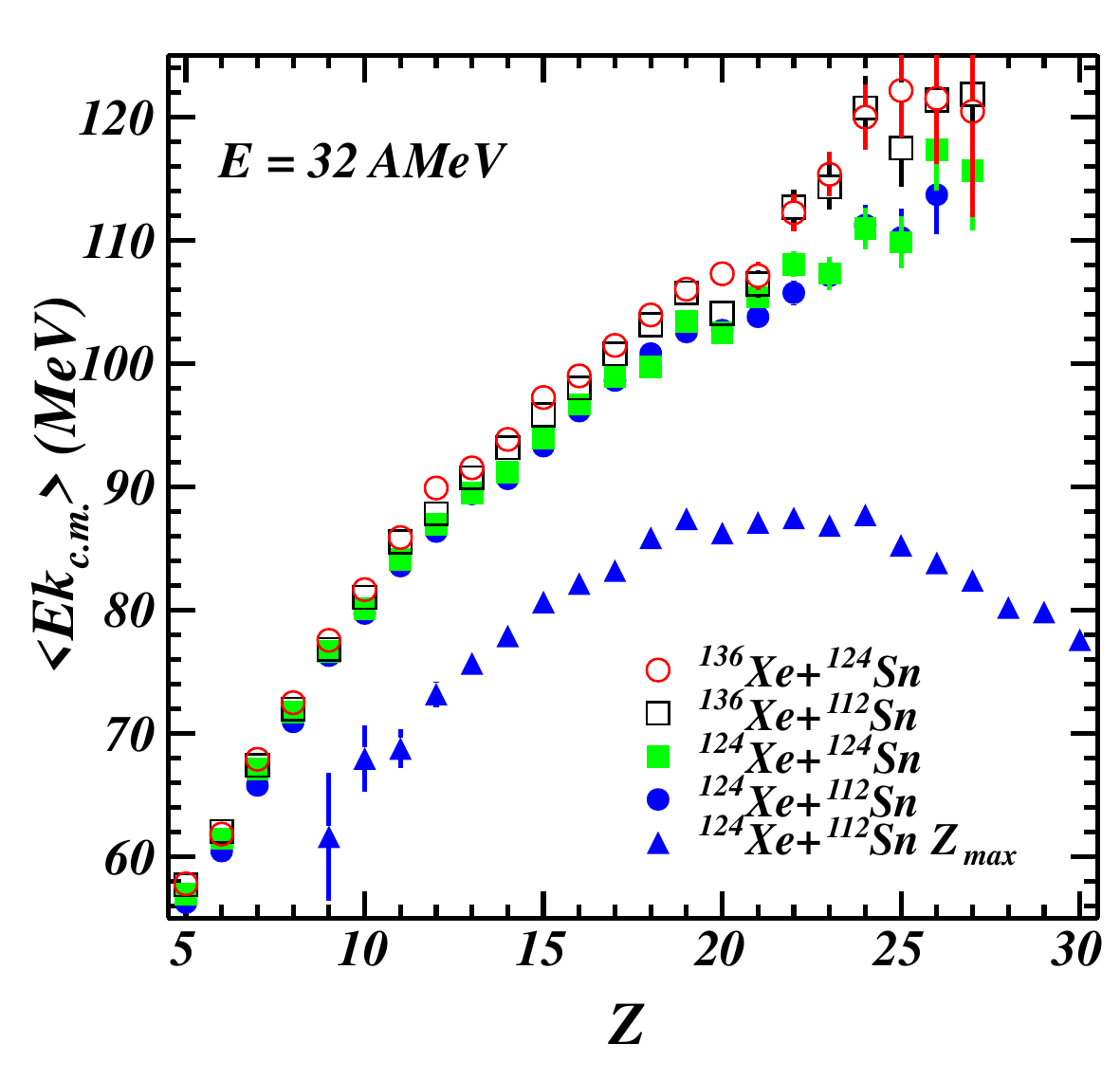}
\includegraphics[width=0.45\textwidth]{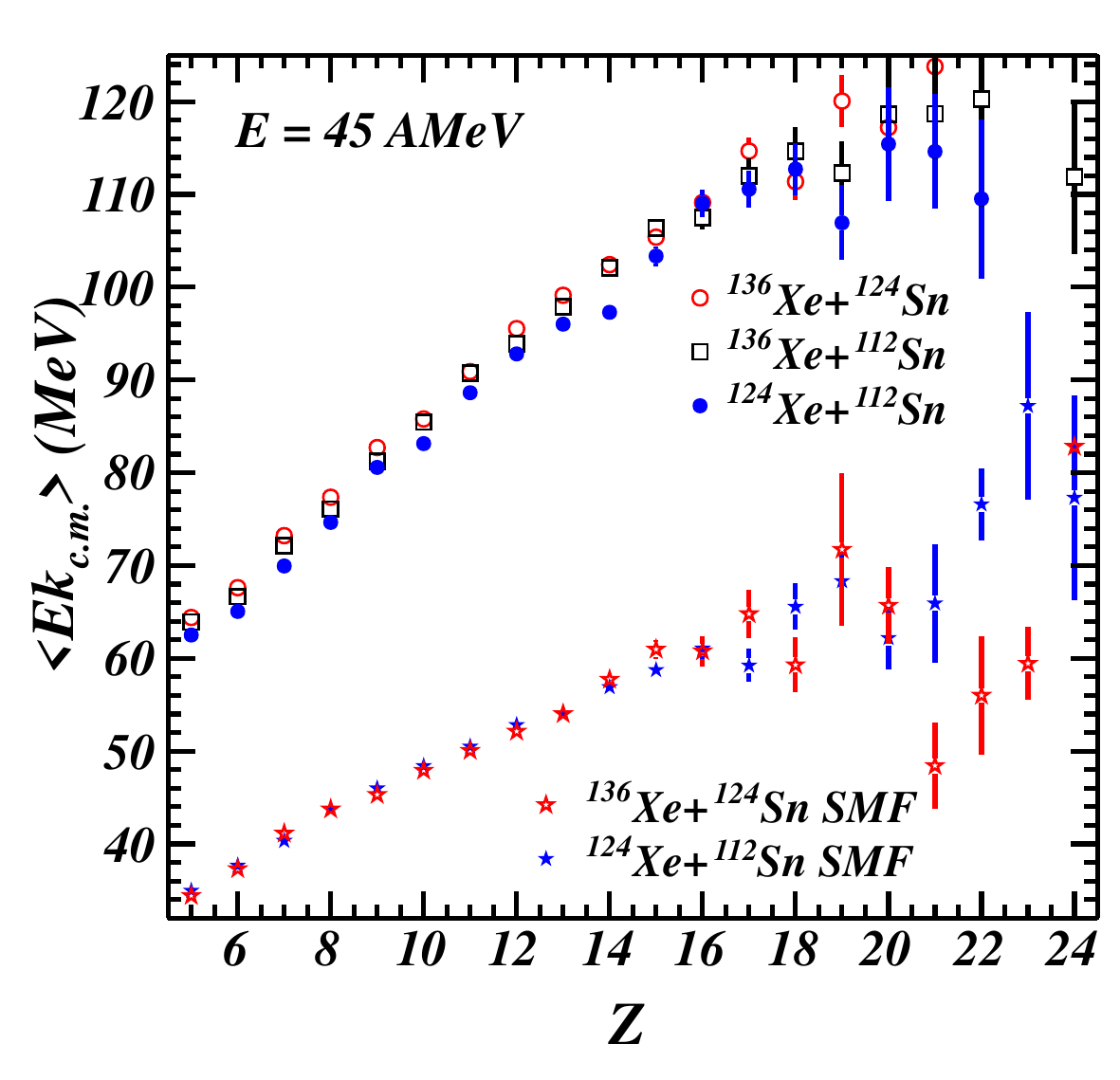}
\caption{(color on line) Average fragment kinetic energy, in the reaction center of mass,
vs the fragment atomic number for the selected events at 32 (left) and 45 
(right) \AM{}. The largest fragment of each partition has been removed.
The average kinetic energy of the largest fragment is displayed for the
\sys{124}{Xe}{112}{Sn} at 32 \AM{}.
Open symbols for \nuc{136}{Xe} projectile, closed symbols 
for \nuc{124}{Xe} projectiles. Circles, squares and triangles: experimental data; 
stars: SMF simulations (cold fragments, filtered). }\label{fig:Ek}
\end{figure*}
The average fragment kinetic energies (in the center of mass) provide 
another piece of information on possible isospin effects. 
Figure~\ref{fig:Ek} displays this observable for the different systems 
at the two incident energies, excluding the largest fragment of each 
partition. The general aspect is the same as that 
previously observed for \sys{129}{Xe}{nat}{Sn}~\cite{I9-Mar97,I39-Hud03,I57-Tab05},
namely an almost linear rise of the average energy when the atomic 
number increases. Conversely the mean energy of the largest fragment 
(see one example in fig.~\ref{fig:Ek}) increases for small Z and then 
decreases. We confirm that,
for a given Z, the mean energy is smaller when it corresponds to the largest 
fragment~\cite{I9-Mar97,I57-Tab05}.
The average kinetic energy for a given element is slightly larger 
(by 2-3 MeV) at the higher incident energy. 

The present data provide
new information: at a given incident energy the mean fragment kinetic energy 
depends on the isospin of the system; the heavier the mass of the system, 
the larger the fragment kinetic energy is.
We have verified that the non-measurement of the masses of the heavy fragments
is not responsible for the observed effect. Indeed the mean measured kinetic
energies, in the laboratory, are larger for the neutron-rich system. The mean 
measured masses for Z=5-8 are also larger. 

The fragment kinetic energy is the superimposition of a disordered thermal
motion and an ordered term. The former term comes from the temperature of 
the fragmenting system at freeze-out; the latter one, radially directed, 
originates from the Coulomb force, which depends on the system 
and fragment charges, and an expansion term proportional to the fragment 
mass~\cite{Hsi94,Dur99,Tam06}. At a given 
available energy the thermal term is expected to be identical for all
systems. The same holds for the Coulomb term as the system charges 
remain very close even after preequilibrium emission (see next section).
Moreover the increase of the fragment multiplicity with the system mass 
should slightly decrease the Coulomb part~\cite{Rad05}, which is not the
observed effect. 
In this context, the increase of the fragment kinetic energy signals the
larger masses of the primary fragments when the system initially contains 
more neutrons. This is true if the expansion energy per nucleon does not
depend on the initial isospin of the system, at a given incident energy.
Isospin differences between systems thus seem to survive preequilibrium
emission~\cite{Lom11}.

\section{\label{sec:ExpSMF}Comparison of data with a Stochastic mean field model}
We used the stochastic mean field (SMF) calculation described in
ref.~\cite{Colon98}. Simulations were performed for collisions corresponding
to all systems (table~\ref{tab:Sys}), 
up to a time equal to 300 fm/$c$. The isoscalar EOS is
soft ($K_{\infty}$=200 MeV), and two parameterisations of the potential part
of the symmetry energy are used~\cite{Gal09}, an asystiff one linearly 
increasing with density while the asysoft form (SKM*) has a maximum around 
normal density. The collision term uses the free nucleon-nucleon cross-section, 
with its isospin, energy and angular dependences\footnote{with an upper
limit of 50 mb, to repress spurious low-energy collisions}. 
This version of the model is 
the same as the one used in ref.~\cite{Gal09}, in which information on the 
asy-stiffness of the EOS
was derived from isospin diffusion in the Ni+Au system at 52 and 74~\AM{}.
There is a great interest in finding additional experimental 
constraints on the density dependence of the symmetry energy in the same 
theoretical framework.

At 45~\AM{} we observe the formation of a single source 
which subsequently breaks into several fragments. 
At 32~\AM{}, conversely to the Brownian One Body
(BOB) calculations~\cite{Gua96}  shown in~\cite{I29-Fra01,I40-Tab03,I57-Tab05},
the systems do not multifragment in head-on collisions.
However it should be noticed that while the BOB calculations were perfomed for
a thermalized source mimicking the composite system at the
moment of maximum compression, here the whole dynamical evolution 
of the system, from the beginning of the reaction,
is simulated by the SMF approach. Hence the impact of pre-equilibrium effects
on the dynamics could be different in the two treatments. 

Fragments are recognized by applying a coalescence procedure to the 
one-body density, connecting nearby cells in which the density is larger 
than a cut-off value, taken equal to $\rho_0/$5 (``liquid phase''). 
We have shown in~\cite{I29-Fra01} that, at 300~fm/$c$, the fragment
multiplicity is independent of the exact value of the cut-off density.  
The remaining early emitted nucleons constitute the ``gas phase''. 
The fragment phase space configuration at 300 fm/$c$ is injected in the 
SIMON code~\cite{Dur92} which performs the secondary decay during the
propagation of all products under the Coulomb field, thus preserving 
space-time correlations. 
Note that the fragment excitation energies at 300 fm/$c$ ($\sim$ 3.3 \AM{}) 
agree well with experimental determinations~\cite{I39-Hud03,I66-Pia08}.
1000 events were run with SMF for each system, then each primary event was 
de-excited 20 times. The cold products were finally 
filtered through a geometrical replica of the INDRA array described in the code 
Panforte~\cite{Nap10}. 
 Because in the present calculations free nucleons (``gas phase'') emitted 
along the dynamical evolution are not taken into account in the subsequent
de-excitation and Coulomb propagation steps, 
the selection on the total detected charge cannot
be used.
Moreover  only fragment properties can be compared to the
experimental values.

\subsection{\label{sec:Mf}Fragment multiplicities}

\begin{figure}
\includegraphics[width=\columnwidth]{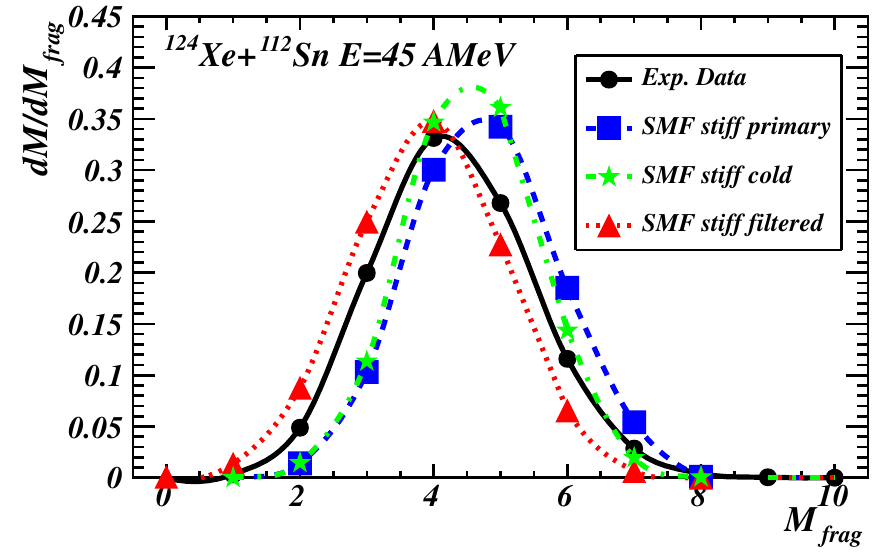}
\caption{(color on line)
Fragment multiplicity distributions, normalised to the number of
events, for \sys{124}{Xe}{112}{Sn} at 45~\AM{}: experimental (circles), 
primary fragments (squares), cold fragments before (stars) and after 
(triangles) filtering. The asystiff EOS is used. Statistical 
errors are smaller than the symbol sizes.} \label{fig:distMf}
\end{figure}

We firstly verified the pertinence of the simulation by comparing calculated
and measured distributions for some observables. We found that the charge
distributions are reasonably reproduced~\cite{GalIWM11}.
We show in Fig.~\ref{fig:distMf} an example of fragment multiplicity 
distributions: for the \sys{124}{Xe}{112}{Sn} at 45~\AM{} the measured
experimental distribution is plotted together with those of primary
fragments, cold fragments, and the filtered distribution. The primary and
cold fragment distributions are very similar, both the width and the average
value of the latter being only slightly smaller than those of the former. 
Let us recall that we consider fragments starting from boron (Z=5).
Two reasons support
this choice in the simulations: first the yield of primary fragments 
is less sensitive to the fragment formation process, which comes from
spinodal decomposition in the SMF model and strongly disfavours the formation
of very small fragments~\cite{Col10}. Second these fragments are essentially
remnants of primary hot fragments, whereas final Li and Be isotopes are also
populated by secondary decay\footnote{The multiplicity of fragments with
Z$\geq$3 is, for all systems, larger after de-excitation than at 300 fm/$c$, 
which is not the case for that of fragments with Z$\geq$5}. 
The filtered distribution shows a good agreement with the experimental one. 

\begin{figure}
\includegraphics[width=0.85\columnwidth]{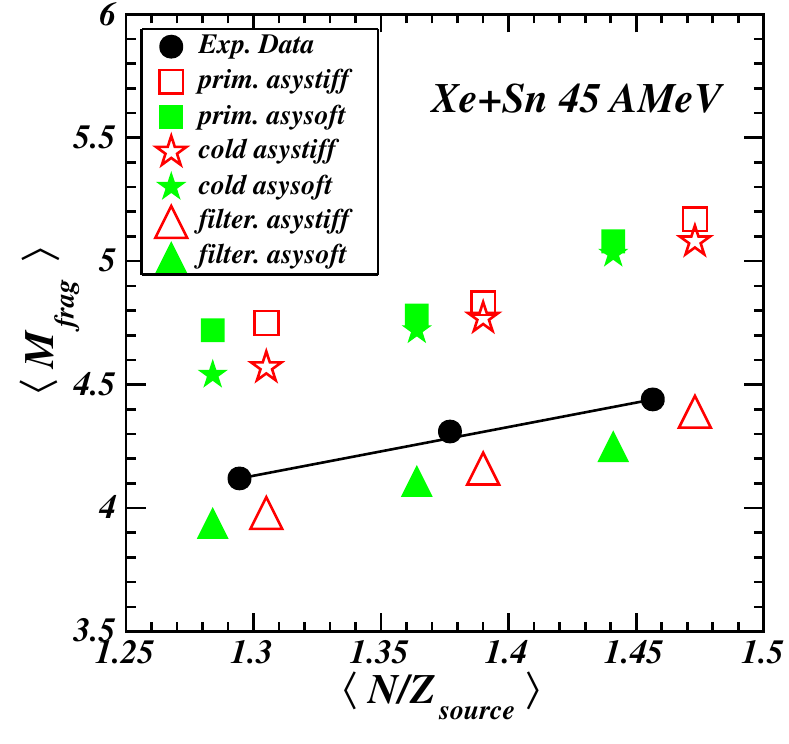}
\caption{(color on line)
Average fragment  multiplicities, for the different systems at 45~\AM{}:
experimental values (circles, the line is to guide the eye),  
calculated values for primary fragments (squares), cold fragments before 
(stars) and after (triangles) filtering, are plotted vs the source N/Z. 
Asystiff EOS: open symbols, asysoft EOS: filled symbols. In all cases
statistical error bars are smaller than the symbol sizes.} \label{fig:MfvsNZ}
\end{figure}

The full set of data is represented on Fig.~\ref{fig:MfvsNZ}. Average
multiplicity values are reported as a function of the N/Z of the sources, 
taken at t=120~fm/$c$ (see next subsection). 
The experimental points are located at the average N/Z of 
the asystiff and asysoft sources. We first observe that the multiplicity of 
primary fragments (squares) increases with the isospin of the multifragmenting 
source, which allows to state that the experimental trend is essentially due to 
the preequilibrium stage  of the collision.
It would support the scaling law of multiplicity with system mass
expected from  a spinodal decomposition. 
The cooling of the fragments only slightly decreases the
multiplicities, and more when the projectile is \nuc{124}{Xe}; indeed
fragments from the less neutron-rich system are expected to evaporate more
charged products, leading to more final fragments with a charge smaller than 5.
Finally for all systems the detection reduces the fragment multiplicity by 
about 0.6 units. These final calculated multiplicities are in good agreement
with the experimental values.
The figure also shows that, although fixed at the
dynamical stage, this observable cannot help in choosing the asystiffness
of the EOS, because at each stage of the calculation the fragment 
multiplicities are located on the same straight line whether the asysoft or
the asystiff EOS was used.

\subsection{\label{sec:Preeq}Preequilibrium emission}
\begin{figure}
\includegraphics[width=0.85\columnwidth]{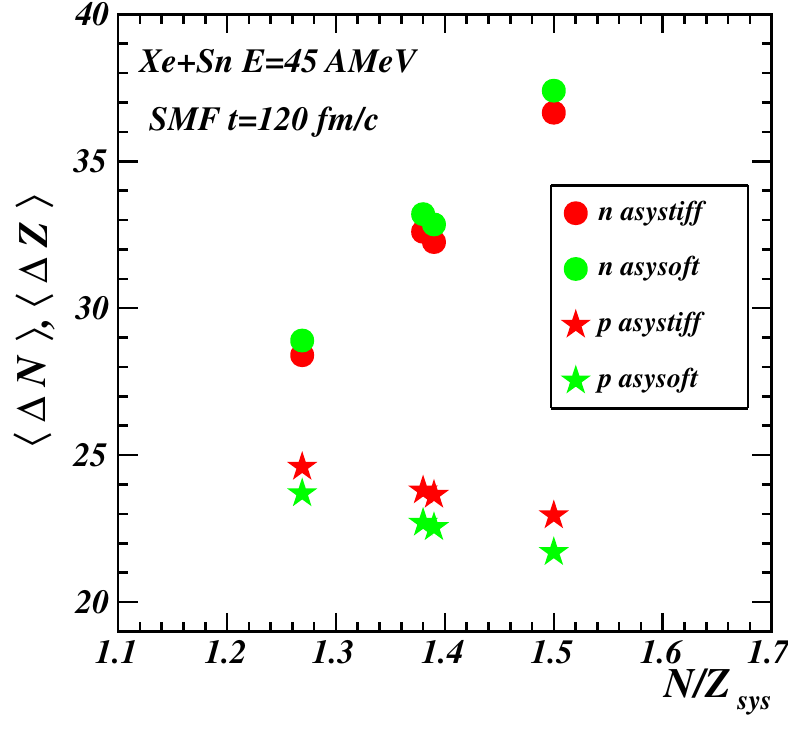}
\caption{(color on line)  Average neutron (circles) and proton (stars) 
preequilibrium emission vs 
the N/Z of the total systems at 45~\AM{}. The two points corresponding to
N/Z=1.38 have been shifted for visibility. Results are shown for the two
asy-EOS.}\label{fig:preeq}
\end{figure}

We define the multifragmenting source as the largest single cluster
recognized by the clusterisation algorithm at t=120 fm/$c$, 
the rest of the system being considered as preequilibrium 
emission. As in ref.~\cite{Col10} we find that the rate of nucleon emission 
in SMF is large between 60 and 120~fm/$c$ to become constant at a small value
afterwards;
this is a general feature of all semi-classical transport models.
The evolution of the number of
preequilibrium nucleons (defined as the difference between numbers of 
neutrons or protons of the system and of the source ) with the N/Z of the 
systems is shown in Fig.~\ref{fig:preeq}, for reactions at 45~\AM{}. 
In all cases about 23\% of the
mass (charge) of the system has been emitted at 120~fm/$c$.
The neutron-richest system has 24 extra-neutrons as compared to the 
neutron-poorest one; we observe that the former system
ejected 8-9 more neutrons, and two protons less,
With an asysoft EOS, there are more neutrons and less protons emitted than
in the asystiff case.

In the model most of that emission occurs during the expansion of the
system, and thus it is a witness of subnormal density EOS~\cite{Col10}; in
that density region the symmetry potential is more repulsive for neutrons,
and more attractive for protons in the asysoft case, which explains the
relative values displayed in Fig.~\ref{fig:preeq}.
Whatever the EOS is, in all cases the neutron to proton ratio of the fast 
emission is more neutron-rich than the system for \sys{136}{Xe}{124}{Sn}, 
and more neutron-poor for  \sys{124}{Xe}{112}{Sn}.
Consequently the isospin content of the multifragmenting 
sources is smaller than that of the total system for
\sys{136}{Xe}{124}{Sn}, and larger for \sys{124}{Xe}{112}{Sn}, 
which reduces the N/Z range  explored in multifragmentation. 
For the intermediate systems, the source is
slightly more neutron-rich than the system with the asystiff EOS, and less
for the asysoft EOS.

\begin{table*}[htb]
\caption{For the 3 systems studied at E/A=45 MeV: mean experimental value of
$Z_{b_5}$; for SMF simulations, and the two asy-EOS: number of protons lost 
at 120~fm/$c$, $Z_{b_5}$ for primary fragments (at 300~fm/$c$), for cold fragments 
and after filtering.
In all cases the numbers between parentheses give the standard deviations of
the corresponding distributions.
}\label{tab:Zb5}
\begin{tabular}{l|c|clccc} \hline
System & Exp.     & \multicolumn{5}{c}{SMF} \\
   & $\langle Z_{b_5}\rangle$& asy & $\langle Z_{pr}\rangle$&$\langle Z_{b_5}^{prim}\rangle$
   &$\langle Z_{b_5}^{cold}\rangle$&$\langle Z_{b_5}^{filt}\rangle$ \\ \hline
\multirow{2}{*}{\sys{124}{Xe}{112}{Sn}}& \multirow{2}{*}{41.9 (8.1)} 
&stiff& 24.60 (1.23) & 63.2 (2.7) & 47.2 (5.06) & 41.6 (8.42) \\  
&&soft& 23.72 (1.23) & 63.9 (2.8) & 47.4 (5.06) & 41.6 (8.52) \\
\multirow{2}{*}{\sys{136}{Xe}{112}{Sn}}& \multirow{2}{*}{44.4 (8.1)} 
&stiff& 23.65 (1.21) & 64.4 (2.8) & 48.9 (5.21) & 42.9 (8.72) \\  
&&soft& 22.55 (1.12) & 65.7 (2.6) & 49.6 (5.17) & 43.4 8.90) \\
\multirow{2}{*}{\sys{136}{Xe}{124}{Sn}}& \multirow{2}{*}{46.2 (8.3)} 
&stiff& 22.95 (1.20) & 65.2 (2.9) & 49.8 (5.48) & 43.4 (8.96) \\  
&&soft& 21.71 (1.13)  & 66.6 (2.7) & 50.6 (5.42) & 44.0 (9.06) \\ \hline
\end{tabular}
\end{table*}

Isolating preequilibrium emission in experiments is a difficult task.
A possible choice is to look at high energy nucleons, 
possibly including nucleons bound in
clusters~\cite{Sap01,Zha08}. We propose as an alternative to 
look at the complementary part of the proton preequilibrium emission, 
the ``liquid phase'', through the value of the total charge bound in fragments, 
$Z_{b_5}=\sum_1^{M_{frag}} Z_{Z\geq5}$. 
Obviously this variable  is only meaningful provided that the 
detected events contain a large part of the total charge of the system, which 
is the case of our selected events. 
We want to test the dependence of $Z_{b_5}$ on the symmetry energy
term implemented in the simulations. 
In this aim we have followed the 
evolution  of $Z_{b_5}$ as a function of the collision stage, for 
the measured systems, and compared the final result of the calculations 
with the experimental data.
We show in Table~\ref{tab:Zb5} 
the  mean values and standard deviations of the experimental and calculated
(primary, and cold before and after filtering) distributions for the different 
systems. We clearly observe that the experimental
distributions are shifted upwards for the neutron-rich systems.
In simulations the mean values of the distributions decrease whereas the
standard deviations increase when going from primary to cold fragments
and to the filtered distributions. The simulated mean values at all stages
are, as in the experiment, larger when the system is more neutron-rich. 
The filtered mean value agrees rather well with the experimental one for
\sys{124}{Xe}{112}{Sn}, while it is more and more underestimated for the two
other systems. We also notice that the filtered distributions are broader
than the measured ones.
We remark that the final calculated $\langle Z_{b_5} \rangle$ does not 
depend on the
asy-stiffness of the EOS for the neutron-poor system whereas it becomes 
larger in the asy-soft case when the system contains more neutrons. 

If we compare the results for the extreme systems, \sys{136}{Xe}{124}{Sn}
and \sys{124}{Xe}{112}{Sn},
the difference between the number of preequilibrium protons is negative 
(fourth column of table~\ref{tab:Zb5} and Fig.~\ref{fig:preeq}). 
Therefore we expect a positive difference, $\Delta Z_{b_5}$, when we turn 
to the values of $Z_{b_5}$, which is
observed in the model for primary fragments at 300~fm/$c$, and for cold
fragments. Note that the difference between the two systems grows larger
during secondary decay, indicating that while it is essentially due 
to the first stage of the collision, part of it comes from the evaporation 
process. The filtering
makes the $\Delta Z_{b_5}$ smaller again. 
At all steps the difference between the two systems is larger for the
asysoft EOS. 
Experimentally we do observe a positive $\Delta Z_{b_5}$, which is larger
than what is obtained with any of the two EOS. The authors of ref~\cite{Col10}
underline that the calculated $\Delta Zb_5$ values depend on the fragment 
formation process; the spinodal decomposition which drives
multifragmentation in SMF strongly disfavours the formation of light
fragments as compared with an Antisymmetrized Molecular Dynamics calculation,
AMD, (see figure 8 of ref.~\cite{Col10}). However in view of the significant 
measured difference between the two systems, we think that $\Delta Z_{b_5}$ 
might be in the future a good observable for constraining the EOS. 

\subsection{Fragment kinetic energies}
 The calculated average kinetic energy of the final fragments at 45~\AM{}
 is represented
in Fig.~\ref{fig:Ek}. As in~\cite{I29-Fra01}, we have added to the average 
kinetic energies a thermal term (3T/2) which is not contained in the
calculation.    
One immediately observes that the model largely
underestimates the measured energies.
The underestimation of the energy of fragments produced through
multifragmentation in central collisions seems to be a general drawback of 
semi-classical transport models. In~\cite{I29-Fra01}, we found 
that in the BOB simulation the fragment kinetic energies were about 20\% 
below the experimental values for \sys{129}{Xe}{nat}{Sn} at 32~\AM{}. 
Several causes of the discrepancy were identified~\cite{I12-Riv98}:
thermal fluctuations were reduced by the large number of test particles,
whereas quantal fluctuations were neglected. Second the finite range of the
force introduced in the model was too large at normal density, creating
surface energy at the detriment of fragment kinetic energy; the initial
expansion energy was largely canceled. 
With the present SMF version the discrepancy between 
calculation and experiment reaches $\sim$40\%. In this model the range 
of the force is correct for all densities; thermal fluctuations are still
underestimated and in  addition the large amount of preequilibrium emission 
considerably reduces the available energy of the fragmenting system. 
Indeed in reference~\cite{Col10} it is shown that the AMD model produces 
less preequilibrium nucleons, and fragments with larger kinetic energies. 
The overestimation of particle emission seems to be a rather general 
problem of semi-classical approaches, as also stressed in~\cite{Lac98}.
However, it is interesting to note that in
all simulations (BOB, SMF, AMD) for Sn+Sn or Xe+Sn collisions between 30 and
50~\AM{} the excitation energy of the fragments is found close to 3~\AM{}.
The lack of available energy in SMF with respect to AMD thus only appears in the
kinetic energy of the fragments. 
 Finally the isospin effect experimentally observed is not predicted by 
 the calculation, which is not surprising if the effect comes from the
expansion energy, which is consumed during the fragment formation.

Coming back to the results displayed in table~\ref{tab:Zb5}, we notice that the
values of $Z_{b_5}$ from the filtered model and the experiment are similar, which 
would imply a correct number of nucleons emitted at 300 fm/$c$.
We should however stress again that we could not apply the exact experimental 
selections to the calculation (completeness of the events and use of a
simple geometrical filter). 

\section{Summary and conclusions}

We have studied central collisions between different isotopes of Xe and Sn 
nuclei, leading to systems differing by their number of neutrons, 
at incident energies of 32 and 45~\AM{}.
Experimentally we highlighted several isospin effects. At a given incident
energy the lcp multiplicity decreases whereas the fragment multiplicity
increases with the N/Z of the system. The stochastic mean field model
developed in Catania well accounts for the fragment partition properties at
45~\AM{},  while at 32~\AM{} the model is not able to describe the 
multifragmentation process.
Comparing model and experimental data, we infer that the increase of the 
fragment multiplicity essentially comes from the dynamical part of the 
reaction, but does not help in constraining the symmetry energy term 
of the EOS. Indeed this seems to be related just to the larger size
of the fragmenting source in the neutron-richer cases.

We also found that on average the fragment kinetic energy grows larger 
when the isospin of the system increases, and we think that this 
observation confirms the existence of an expansion energy and shows that the
primary fragments keep some sign of the neutron-richness of the initial system. 
The model fails in reproducing the fragment kinetic properties, one of the 
reasons being that too much energy is removed from the system by
preequilibrium emission.

Finally we propose to use the charge bound in fragments as
an alternate variable to get information on lcp preequilibrium emission. 
We experimentally observe 
a sizeable difference between the values of $Z_{b_5}$ measured for
\sys{136}{Xe}{124}{Sn} and \sys{124}{Xe}{112}{Sn}, which makes it a
promising observable to constrain the symmetry energy. The calculated values 
of this difference depend on the asystiffness of the EOS, but are smaller
than the measured one both in the asysoft and asystiff case. 

In prospective further developments of stochastic transport models are in 
progress. Phase space fluctuations are introduced by a stochastic treatment
of the nucleon-nucleon collision integral. Once a two-body collision occurs, two
clouds of test particles, which simulate the wave-packet extension of the
colliding nucleons, are moved simultaneously to new phase-space locations.
This procedure, that corresponds to a numerical implementation of the 
Boltzmann-Langevin equation, has been proven to give the correct amplitude 
of equilibrium thermal fluctuations~\cite{Rizz08}.
First tests with this model show an improvement of the description 
of the kinematical and partition properties of fragmenting systems in 
dissipative collisions around the Fermi energy~\cite{*[{}] [{ and paper in
preparation.}] NapIWM11}.  


%

\end{document}